\documentclass[preprint,review,10pt]{elsarticle}

\usepackage{amssymb}

 \usepackage{amstext}
 \usepackage{multirow}
 \usepackage{graphicx}

 \usepackage{amsmath}
 \usepackage{algorithmic}
 \usepackage{algorithm}
 \usepackage{bigstrut}
\usepackage{pifont}
\usepackage{natbib}
\usepackage{geometry}
\usepackage{fleqn}
\usepackage{txfonts}
\usepackage{amsfonts}

 \usepackage{amssymb}
\usepackage{picins}

\begin{document}

\begin{frontmatter}



\title{Low Complexity Joint Estimation of Synchronization Impairments in
Sparse Channel for MIMO-OFDM System\\ (\textit {under review in AEU -
International Journal of Electronics and Communications (Elsevier)
(paper id-AEUE-D-12-00625))}}
\author[]{Renu Jose\corref{cor1}}
\ead{renujose@ece.iisc.ernet.in}
\author[]{~Sooraj K. Ambat}
 \ead{sooraj@ece.iisc.ernet.in}
\author[]{~K.V.S. ̃Hari}
\ead{hari@ece.iisc.ernet.in}
\cortext[cor1]{Corresponding author, Ph.+91 9483709828; Fax:+91 80
23600563}

\address{Statistical Signal Processing Lab, Department
of Electrical Communication Engineering, Indian Institute of Science, Bangalore-560012,
India.\vspace{-1cm}}

\begin{abstract}
 Low complexity joint estimation of synchronization
impairments and channel in a single-user
 MIMO-OFDM system 
is presented in this letter. Based on a system model that takes into account
the effects of
synchronization impairments such as carrier frequency offset, sampling frequency offset,
and symbol timing error, and channel, a Maximum Likelihood (ML) algorithm for the joint
estimation is proposed. To reduce the complexity of ML grid search, the
number of received signal samples used for estimation need to be reduced. 
The conventional channel estimation methods using Least-Squares (LS) fail for
the reduced sample under-determined system, which results in poor performance
of the joint estimator. 
The proposed ML algorithm uses Compressed Sensing (CS) based channel estimation
method in a sparse fading scenario, where the received samples used for
estimation are less than that required for an LS based estimation. The
performance of the estimation method is studied through
numerical simulations, and it is observed that CS based joint estimator 
performs better than LS based joint estimator.
\end{abstract}
\begin{keyword}
 MIMO\sep OFDM\sep Synchronization\sep Channel Estimation\sep Sparse
Channel\sep Compressed Sensing.
\end{keyword}
\end{frontmatter}
\section{Introduction}
\label{}
\vspace{-.3cm}
Multiple Input Multiple Output-Orthogonal Frequency Division
Multiplexing (MIMO-OFDM) system, the preferred solution for the next generation
wireless technologies, is very sensitive to synchronization impairments
such as Carrier Frequency Offset (CFO), Sampling Frequency Offset (SFO) and
Symbol Timing Error (STE)
~\cite{morelli_tutorial}-\cite{renu2}. 
In this letter, we propose a low complexity Maximum Likelihood (ML) algorithm
for the joint estimation of synchronization impairments and channel using 
Compressed
Sensing (CS) technique, in a sparse
fading scenario, where the received samples used for estimation are less than
that required for a Least Squares
(LS) based estimation.
\vspace{-.5cm} 
\section{System Model}\label{system_model}
\vspace{-.3cm}
\par
Consider a MIMO-OFDM system with $N_T$ transmit antennas and $N_R$ receive
antennas using Quaternary Phase Shift Keying (QPSK) modulation and $N$
subcarriers per antenna. Let $T$ be the
sampling time at
the transmitter and $f_c$ be the carrier frequency. We define the normalized CFO
as
$\epsilon=\Delta f_cNT$, the normalized SFO as $\eta=\Delta T/T$, and
the normalized STE as $\theta$, where $\Delta f_c $ is the net CFO in the
received signal and $\Delta T$ is the difference between the sampling time at
the receiver and the transmitter~\cite{renu2}. Let $\mathbf{X}$ be the block diagonal
matrix with each diagonal matrix having the signal vector transmitted from
each transmit antenna. Also, let $\mathbf{h}$ be the column vector representing
the MIMO channel with $L_m$ as the maximum length of channel between any
transmit and receive antenna pair. The signal vector at the receiver side is derived
in~\cite{renu2} as,
\begin{align}\label{datamodel1}
&\mathbf{r}=\mathbf{A}_1(\epsilon,\eta,\theta)\mathbf{h}+\mathbf{w}\\
&\mathrm{where}
\hspace{.1cm}\mathbf{A}_1(\epsilon,\eta,\theta)=\mathbf{I}_{N_R}\otimes
(\mathbf{D}(\epsilon,\eta)\mathbf{F}_1(\eta)\mathbf{G}(\theta)\mathbf{X
(\mathbf{ I } } _ { N_T } \otimes
\mathbf{F}_2))\nonumber\\
&\mathbf{D}(\varepsilon,\eta)=
{diag}[1,\exp(j2\pi\varepsilon(1+{\eta})/N),\hdots
,\exp(j2\pi\varepsilon(1+ { \eta } )
(N-1)/N)]\nonumber\\
&\mathbf{G}(\theta)=
{diag}[1,\exp(-j2\pi\theta/N),\hdots,\exp(-j2\pi(N-1)\theta/N)],\nonumber\\
&[\mathbf{F}_1(\eta)]_{n,k}=\frac{\exp(j2\pi k(n(1+{\eta}))/N)}{N},
[\mathbf{F}_2]_{k,l}=\exp(-j2\pi lk/N),\nonumber
\end{align} 
\let\thefootnote\relax\footnotetext{
{Notations:}
Upper case bold letters denote matrices and lower case bold letters denote
column vectors. 
$\hat{\mathbf{A}}$ denotes the estimate of $\mathbf{A}$.
  $\mathbf{I}_M$ denotes an $M\times M$ identity matrix. 
identity matrix. 
 $\mathbf{A}^H$ and $\mathbf{A}^\dag$ denote
conjugate
transpose, and pseudo-inverse of $\mathbf{A}$, respectively.
$[\mathbf{A}]_{m,n}$ denotes the $(m, n)^\mathrm{th}$ element of $\mathbf{A}$.
 $\otimes$ represents Kronecker product.
 ${diag}[\mathbf{x}]$ represents a diagonal matrix having the elements
of $\mathbf{x}$ as
diagonal elements.
$\mathrm{Tr}\mathbf{(A)}$ represents trace of $\mathbf{A}$.
 Calligraphic letter $\mathcal{T}$ denotes set and $\mathcal{T}^{c}$
denotes set complement.
$\mathbf{A}_{\cal{T}}$($\mathbf{A}_{(\cal{T})}$)  denotes the column (row)
sub-matrix of $\mathbf{A}$ formed by the columns (rows) of $\mathbf{A}$ listed
in the set $\cal{T}$.
 }
with $n, k=0,1,\hdots,N-1$, and $l=0,1,\hdots,L_m-1$. $\mathbf{w}$ is the
 additive circular Gaussian noise vector with mean zero and variance
$\sigma_{\mathbf{w}}^2$. Let $\theta_{\mathrm{max}}$
 denote the maximum STE. Then the system model in (\ref{datamodel1}) can be
re-written as,
\begin{align}\label{datamodel2}
\mathbf{r}&=\mathbf{A}_2(\epsilon,\eta) \mathbf{{h}}_\theta +\mathbf{w}\\
\mathrm{where}\hspace{.2cm}
&\mathbf{A}_2=\mathbf{I}_{N_R}\otimes
(\mathbf{D(\epsilon,\eta)F}_1(\eta)\mathbf{X(\mathbf{I}}_{N_T}\otimes
\mathbf{F}_{2\theta_{\mathrm{max}}})),\nonumber\\
\mathrm{and}\hspace{.2cm}&[\mathbf{F}_{2\theta_{\mathrm{max}}}]_{k,l}
=\exp(-j2\pi lk/N),\nonumber
\end{align}
with $k=0,1,\hdots,N-1$, $l=0,1,\hdots,(L_m+\theta_{\mathrm{max}}-1)$, and
$\mathbf{{h}}_\theta$ being the STE embedded MIMO channel as given
in~\cite{renu2}. 
\section{ML Algorithm for Joint Estimation}
The ML cost function~\cite{renu2} of the parameters
$\epsilon,\theta,\eta,$ and $\mathbf{h}$, obtained from
(\ref{datamodel1}) is, 
\begin{equation}\label{costfunction}
\arg\min_{\epsilon,\eta,\theta,\mathbf{h}}\mathrm{J}(\epsilon,\eta,\theta,
\mathbf{ h }
|\mathbf{r})=\arg\min_{\epsilon,\eta,\theta,\mathbf{h}}(\mathbf{r}-\mathbf{A}_1
\mathbf{h})^{H}(\mathbf{r}-\mathbf{A}_1\mathbf{h}).
\end{equation}

The multi-dimensional minimization in (\ref{costfunction}) gives the estimate of
the parameters ${\epsilon},{\theta},{\eta}$, and $\mathbf{h}$.
Given the estimate of channel, $\hat{\mathbf{h}}$ and
$\hat{\mathbf{h}}_\mathrm{\theta}$, and using the system models in
(\ref{datamodel1}) and (\ref{datamodel2}), the optimization problem in
(\ref{costfunction}) reduces to a two-dimensional and one-dimensional
minimization problem respectively as,
\begin{equation}\label{costfunction1}
\hspace{-.6cm}
[\hat{\epsilon},\hat{\eta}]=\arg\min_{\epsilon,\eta}(\mathbf{r
}-\mathbf{A}_2
\hat{\mathbf{h}}_{\theta})^{H}(\mathbf{r}-\mathbf{A}_2\hat{\mathbf{h}}_{\theta}
)=\arg\min_ { \epsilon, \eta ,
}\mathrm{J}_1(\epsilon,\eta|\mathbf{r},\hat{\mathbf{h}}_{\theta}),
\end{equation}
\begin{equation}\label{costfunction2}
\hspace{-.6cm}
[\hat{\theta}]=\arg\min_{\theta}(\mathbf{r
}-\mathbf{A}_1
\hat{\mathbf{h}})^{H}(\mathbf{r}-\mathbf{A}_1\hat{\mathbf{h}})=\arg\min_{
\theta
}\mathrm{J}_2(\theta|\mathbf{r},\hat{\epsilon},\hat{\eta},\hat{\mathbf{h}}).
\end{equation}

For the above ML algorithm to have a unique
solution with the LS estimate of the channel, the number of
received signal samples used for estimation must at least be equal to the number
of unknown channel coefficients, i.e., $MN_R\geq L_mN_TN_R$. To have a low
complexity joint estimation at the receiver we need to reduce the received
samples used for estimation, where the ML algorithm using LS channel estimation
(MLLS) fails. Hence we propose an ML algorithm using CS technique which performs
better than MLLS for an under-determined MIMO-OFDM system
in sparse fading channel. 
\subsection{CS based channel estimation}
\begin{algorithm}
\small
\label{SPAlgo}
\caption{Sparse channel estimation using SP Algorithm}
{\bf Inputs:} $\mathbf{A}$, $\mathbf{r}$, and $K$ 
\label{OMPAlgo}
\begin{algorithmic}[1]
  \STATE $\mathbf{A}=\mathbf{AC}$;\hfill $\star$  Normalize columns of
$\mathbf{A}$ using
diagonal matrix $\mathbf{C}$:\\
       \STATE Initialization: $k = 0$, $\mathcal{T}_0 =\varnothing$, $\mathbf
e_0 =
\mathbf r$;
    \REPEAT
  \STATE $k = k + 1$;
    \STATE $\tilde{\mathcal{T}} =  \mathcal{T}_{k-1} \cup$ \{indices of
$K$-highest magnitude components of  $\mathbf A^H \mathbf e_{k-1}$\}
     \STATE $\mathbf v_{\tilde{\mathcal{T}}} = \mathbf
A_{\tilde{\mathcal{T}}}^\dagger \mathbf r$, $\mathbf v_{\tilde{\mathcal{T}}^c} =
\mathbf 0$ ;
    \STATE $\mathcal{T}_k = $ indices of $K$-highest magnitude components of
$\mathbf v$;
    \STATE $\mathbf e_k = \mathbf r - \mathbf A_{\mathcal{T}_k} \mathbf
A_{\mathcal{T}_k}^\dag  \mathbf r$;
  \UNTIL{($\|\mathbf e_k\|_2 \ge  \|\mathbf e_{k-1}\|_2$)}
\STATE $\mathcal{T}_k =\mathcal{T}_{k-1}$;
\STATE $\mathbf{\hat{h}}_{\mathcal{T}_k} = \mathbf A_{\mathcal{T}_k}^ \dag
\mathbf r$,  $\mathbf{\hat{h}}_{\mathcal{T}_k^c} = \mathbf 0$;
\end{algorithmic}
 {\bf Output:} $\mathcal{T}_k$ ,
$\hat{\mathbf{h}}^{(\mathrm{SP})}=\mathbf{C\hat{h}}$
\end{algorithm}
\begin{algorithm}
\small
\label{MLAlgo}
\caption{MLSP}
\flushleft{\bf Inputs:} $\mathbf{r}_u$, $\mathcal{T}$,
$[\theta_{\mathrm{min}},\theta_{\mathrm{max}},\theta_{\mathrm{grid}}]$,$[
\epsilon_{\mathrm{min}},\epsilon_{\mathrm{max}},
\epsilon_{\mathrm{grid}}]$,$[\eta_{\mathrm{min}},\eta_{\mathrm{max}},\eta_{
\mathrm{grid}}]$
\label{OMPAlgo}
 \begin{algorithmic}[1]
\vspace{-.3cm}
    \FOR{$j=$
$\epsilon_{\mathrm{min}}:\epsilon_{\mathrm{grid}}:\epsilon_{\mathrm{max}}$}
      \FOR{$k=$ $\eta_{\mathrm{min}}:\eta_{\mathrm{grid}}:\eta_{\mathrm{max}}$}
	\STATE Construct $\mathbf{A}_{2u}(j,k)$;\hfill
$\star$ using
(\ref{datamodel3})
	\STATE Obtain $\hat{\mathbf{h}}_{\theta_{j,k}}^{(\mathrm{SP})}$;\hfill
$\star$ using
Algorithm $1$
	\STATE Evaluate
$\mathrm{J}_1\left(j,k|\mathbf{r}_u,\hat{\mathbf{h}}_{\theta_{j,k}}^{(\mathrm{SP
} ) }\right )$;\hfill
$\star$ using
(\ref{costfunction1})
      \ENDFOR
    \ENDFOR
\STATE $\displaystyle
[\hat{\epsilon}_{\mathrm{MLSP}},\hat{\eta}_{\mathrm{MLSP}}]=\arg \min_{j,k}
\mathrm{J}_1\left(j,k|\mathbf{r}_u,\hat{\mathbf{h}}_{\theta_{j,k}}^{(\mathrm{SP}
) }\right)$;
\FOR{$i=$ $\theta_{\mathrm{min}}:\theta_{\mathrm{grid}}:\theta_{\mathrm{max}}$}
\STATE Construct
$\mathbf{A}_{1u}(i,\hat{\epsilon}_{\mathrm{MLSP}},\hat{\eta}_{\mathrm{MLSP}}
)$;\hfill
$\star$ using
(\ref{datamodel3})
\STATE Obtain $\hat{\mathbf{h}}_{i}^{(\mathrm{SP})}$;\hfill $\star$ using
Algorithm $1$
	\STATE Evaluate
$\mathrm{J}_2\left(i|\mathbf{r}_u,\hat{\epsilon}_{\mathrm{MLSP}},\hat{\eta}_{
\mathrm { MLSP}},\hat{\mathbf{h}}_{i}^{(\mathrm{SP})}\right)$;\hfill
$\star$ using
(\ref{costfunction2})
      \ENDFOR  
\STATE $\displaystyle
[\hat{\theta}_{\mathrm{MLSP}}]=\arg\min_{i}\mathrm{J}_2\left(i|\mathbf{r}_u,\hat
{ \epsilon}_{\mathrm{MLSP}},\hat{\eta}_{\mathrm{MLSP}},\hat{\mathbf{h}}_{i}^{
(\mathrm{SP})}\right)$;
\STATE
$\hat{\mathbf{h}}_{\mathrm{MLSP}}=\hat{\mathbf{h}}_{{\hat{\theta}_{\mathrm{MLSP}
}}}^{(\mathrm{SP})}$
\end{algorithmic}
{\bf Output:}
$[\hat{\theta}_{\mathrm{MLSP}},\hat{\epsilon}_{\mathrm{MLSP}},\hat{\eta}_{
\mathrm{MLSP}},\hat{\mathbf{h}}_{\mathrm{MLSP}}]$ 
 \end{algorithm}
\normalsize
CS is a novel technique where a parameter that is sparse in
a transform domain can be estimated with
fewer samples than usually required~\cite{Donoho}~\cite{Candes2}. 
The application of CS is to recover the $K-$sparse channel (A channel
 is said to be $K$-sparse if
 it contains at most $K$ non-zero coefficients) from
$MN_R$ received signal samples, where $MN_R < L_mN_TN_R$. Using
 (\ref{datamodel1}) and (\ref{datamodel2}), 
\begin{align}\label{datamodel3}
 \mathbf{r}_u=\mathcal{F}(\mathbf{r})=\mathbf{A}_{1u}(\epsilon,\eta,
\theta)\mathbf { h } +\mathbf { w }
_u=\mathbf{A}_{2u}(\epsilon,\eta)\mathbf{h}_{\theta}+\mathbf{w}_u,
\end{align}
 where $\mathcal{F}(\mathbf{r})$ is the operator which randomly selects 
$M$ samples from each receive antenna given in
$\mathbf{r}$. Also, $\mathbf{A}_{1u}$=${\mathbf{A}_1}_{(\mathcal{T})}$ and
$\mathbf{A}_{2u}$=${\mathbf{A}_2}_{(\mathcal{T})}$ where $\mathcal{T}$
contains the indices of the $MN_R$ samples selected from $\mathbf{r}$.
 In CS framework, $\mathbf{r}_u$ is called the observation vector and
$\mathbf{A}$,
 which represents either $\mathbf{A}_{1u}$ or $\mathbf{A}_{2u}$, is called the measurement matrix.
\par
In this letter, we use Subspace Pursuit (SP) algorithm~\cite{SP} which is a
popular greedy algorithm used in CS. In each iteration, SP identifies a
$K$-dimensional space that
reduces the reconstruction error of the sparse channel $\mathbf{h}$. 
 The steps involved are given in Algorithm $1$.
 It has been shown theoretically that SP algorithm converges in finite number of
steps \cite{SP}.
\subsection{ML algorithm using SP channel estimation (MLSP)}
To obtain MLSP, the estimate of $\mathbf{h}$ using SP, denoted as
$\hat{\mathbf{h}}_{\theta}
^{(\mathrm{SP})}$ and $\hat{\mathbf{h}}^{({\mathrm{SP}})}$, obtained from
Algorithm 1 are used to 
 rewrite the cost function in (\ref{costfunction1}) and (\ref{costfunction2})
as,
$\displaystyle
\mathrm{J}_1\left(\epsilon,\eta|\mathbf{r}_{u},\hat{\mathbf{h}}_{\theta}
^{(\mathrm{SP})}\right)$ and $\displaystyle
\mathrm{J}_2\left(\theta|\mathbf{r}_{u},\hat{\epsilon},\hat{\eta},
\hat{\mathbf{h}}^{(\mathrm{SP})}\right)$, respectively. 
The steps involved in MLSP are given in Algorithm
$2$.\\
\it{Remarks:}\rm The computational complexity of LS based estimation
in MLLS is approximately $\mathcal{O}((L_mN_TN_R)^3)$, whereas that of
SP based estimation in MLSP is approximately
$\mathcal{O}(MN_R^2N_TL_mK)$~\cite{SP} which is lesser.
\vspace{-.55cm}
\section{Simulation Results and Discussions}\label{simulation}
\vspace{-.35cm}
 We considered a $2\times 2$ MIMO-OFDM system having $N=128$ subcarriers for
each
transmitter with $20$ MHz signal
bandwidth.  The channel coefficients are modeled as 
circular complex-valued Gaussian random variable having unit variance, and
 uniform power delay profile with $L_m$=$26$ and sparsity level, $K$=$5$.
 Also, the transmitted symbols belong to QPSK constellation with unit magnitude. 
We considered the training blocks having a Cyclic Prefix (CP) of length $32$.
The
condition
$(L_m+\theta_{\mathrm{max}})$ less
than length of CP~\cite{renu2} results in $\theta_{\mathrm{max}}$=$5$ and
$|\theta|<5$.
 The range of normalized CFO used for grid search is
$|\epsilon|<0.4$ with a resolution of $10^{-2}$ and that of normalized SFO
 is $|\eta|<5\times10^{-3}$ with a resolution of $10^{-4}$.
The actual values of the impairments, $\epsilon$, $\eta$, and $\theta$ used in
the simulations are $0.102$, $101$ $\mathrm{ppm}$, and $2$,
respectively.  
\begin{figure}[htb]
\centering{
\resizebox{3in}{2in}{
\includegraphics[]{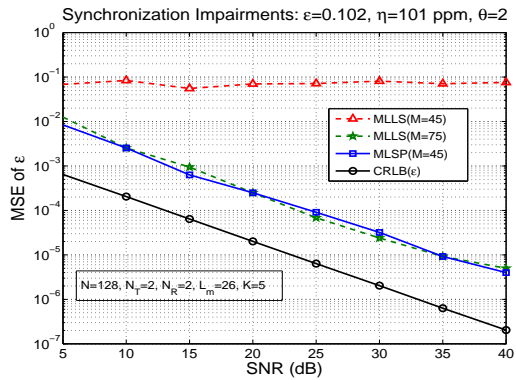}}}
\caption{CRLB and MSE for the estimation of
CFO.}
\label{MSECFO}
\end{figure}
\begin{figure}[htb]
\centering{
\resizebox{3in}{2in}{
\includegraphics[]{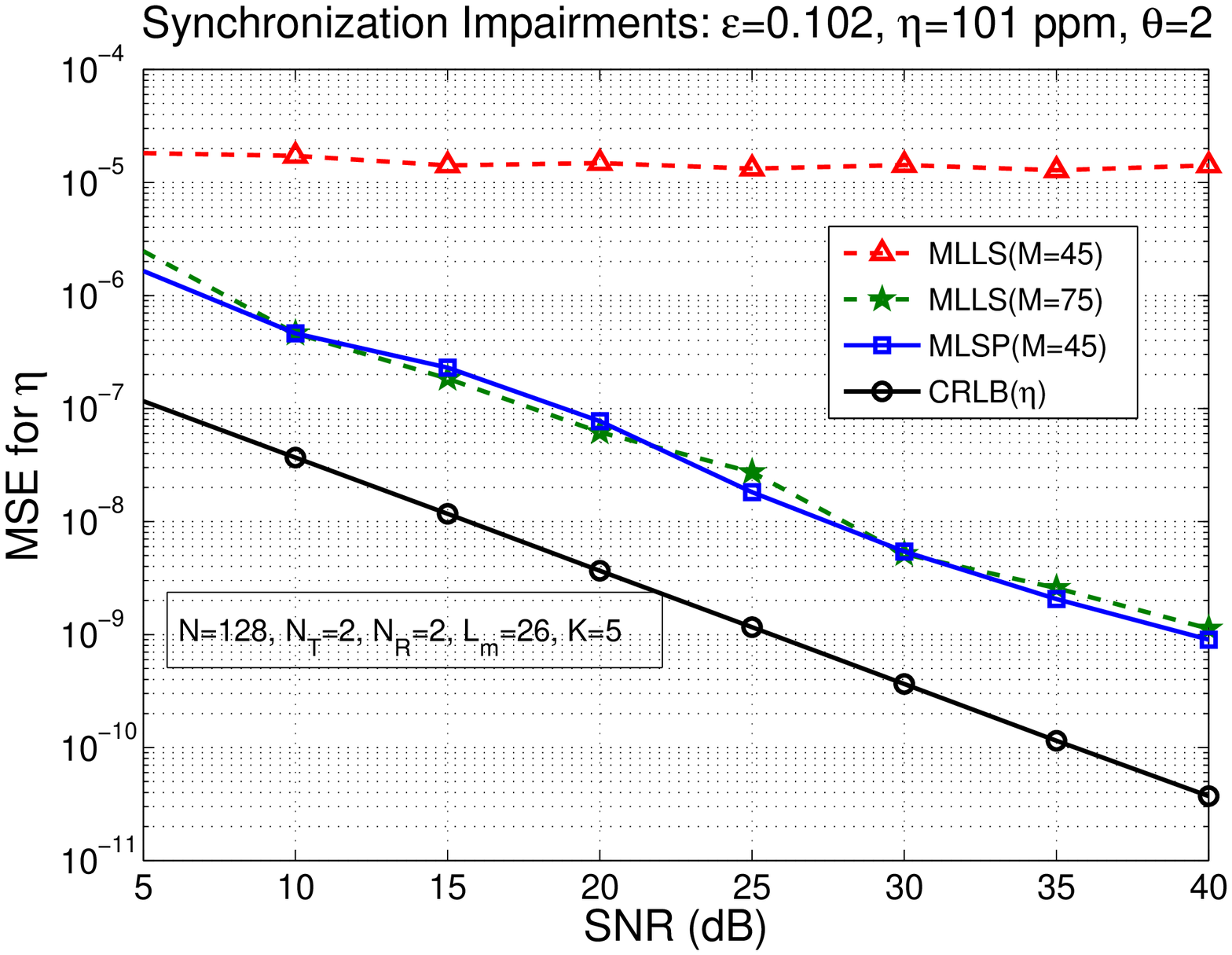}}}
\caption{CRLB and MSE for the estimation of SFO.}%
\label{MSESFO}
\end{figure}
\begin{figure}[htb]
\centering{
\resizebox{3in}{2in}{
\includegraphics[]{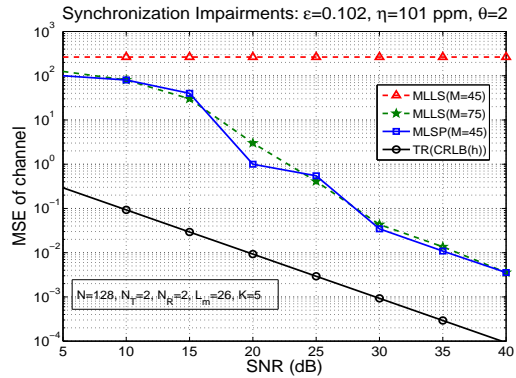}}}
\caption{Tr(CRLB(h)) and MSE for the estimation of channel.}
\label{MSEH}
\end{figure}

\begin{figure}[htb]
\centering{
\resizebox{3in}{2in}{
\includegraphics[]{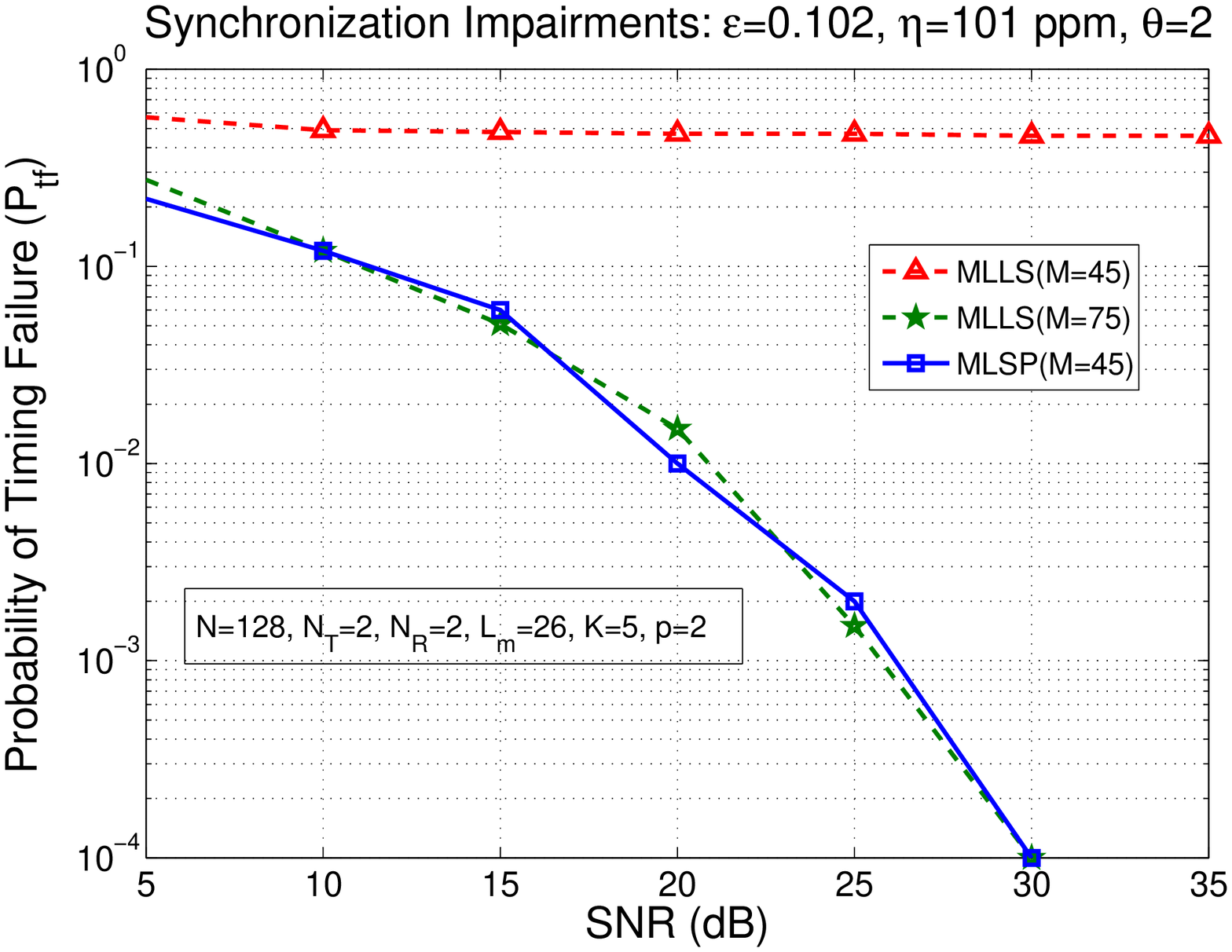}}}
\caption{Probability of Timing Failure as a function of
SNR(dB).}
\label{Ptf_cs}
\end{figure}
The Mean Square Error (MSE) values of the estimated parameters, using MLLS
 and MLSP, are calculated and are plotted in log-scale against SNR(dB),
 together with Cram\'{e}r-Rao Lower Bound (CRLB) of the
parameters~\cite{renu2}, in Fig(\ref{MSECFO}).- Fig(\ref{MSEH}). MLLS is
simulated using MLSP algorithm given in Algorithm 2 by replacing the SP
estimate of channel obtained in step 4 and step 5 using LS estimate of the
channel.
 It is found from Fig(\ref{MSECFO}).- Fig(\ref{MSEH}). that the MSE plots
of MLLS for the estimation
 of CFO, SFO, and channel for $M$=$45$ fail, due to the poor
performance of LS
 based estimation in under-determined system. Also, the MSE plots of MLSP
 for the estimation of CFO, SFO, and channel follow $\mathrm{CRLB}(\epsilon)$, $\mathrm{CRLB}(\eta)$,
 and $\mathrm{Tr}(\mathrm{CRLB}(\mathbf{h}))$~\cite{renu2}, respectively, but with a
performance degradation of around $12$ dB, $13$ dB, and $15$ dB SNRs,
 respectively, at high SNR. The Probability of Timing Failure~\cite{renu2} for
 the estimation of $\theta$,
 defined as $\mathrm{P_{tf}}(p)=\mathrm{Pr}\left[|\hat{\theta}-\theta|\geqslant p\right]$,
 is calculated for $p$=$2$ and is plotted in Fig(\ref{Ptf_cs}). for MLLS and
MLSP, respectively. As in
 the cases of CFO, SFO, and channel, MLSP performs better than MLLS for the
estimation of STE also. It is observed from the figures that, to have a
comparable performance with MLSP using $90$ samples ($M$=$45$), MLLS requires at
least
$150$ samples ($M$=$75$), which shows the difference in computational
complexity.
\vspace{-.7cm}
\section{Conclusion}\label{Conclusion}
\vspace{-.3cm}
 In this letter, we presented a low complexity ML joint estimation algorithm
for single-user MIMO-OFDM system,
where the received samples used for estimation are less than that required for
an LS based ML estimation, MLLS. An ML algorithm for the joint estimation
of synchronization impairments and channel using CS based technique, MLSP, is
proposed. 
 It is found from the simulations that MLSP performs better than MLLS for the joint
 estimation of CFO, SFO, STE, and channel.

\end{document}